\documentclass[twocolumn]{aastex63}
\def\apjs{Ap.J.Supp.}

\def\apjl{Ap.J.Lett.}
\def\aap{A\&A}

\def\gr{$\gamma$-ray}

\usepackage{graphicx}
\shorttitle{Sensitivity of CTA to strong IGMF}
\shortauthors{Korochkin et al.}

\begin{document}
\title{Sensitivity reach of gamma-ray measurements for strong cosmological magnetic fields}

\author{Alexander Korochkin}
\affiliation{INR RAS, 60 October anniversary prosp. 7a, Moscow, Russia}
\affiliation{Université de Paris, CNRS, Astroparticule et Cosmologie, F-75013 Paris, France}
\affiliation{Novosibirsk State University, Pirogova 2, Novosibirsk, 630090 Russia}

\author{Oleg Kalashev}
\affiliation{INR RAS, 60 October anniversary prosp. 7a, Moscow, Russia}
\affiliation{Novosibirsk State University, Pirogova 2, Novosibirsk, 630090 Russia}
\affiliation{Moscow Institute for Physics and Technology, 9 Institutskiy per., Dolgoprudny, Moscow Region, 141701 Russia}

\author{Andrii Neronov}
\affiliation{Université de Paris, CNRS, Astroparticule et Cosmologie, F-75013 Paris, France}
\affiliation{Astronomy Department, University of Geneva, Ch. d'Ecogia 16, 1290, Versoix, Switzerland}

\author{Dmitry Semikoz}
\affiliation{Université de Paris, CNRS, Astroparticule et Cosmologie, F-75013 Paris, France}

\begin{abstract} 
    A primordial magnetic field with the strength in the 1-10 pG range can resolve the tension between different measurements of the Hubble constant and provide an explanation for the excess opacity in the 21 cm line at redshift $15<z<20$, if it is present during the recombination and reionization epochs. This field can also survive in the voids of the large-scale Structure in the present day universe. We study the sensitivity reach of the gamma-ray technique for measurement of such relatively strong cosmological magnetic field  using deep exposure(s) of the nearest hard spectrum blazar(s) with CTA telescopes. We show that the gamma-ray measurement method can sense the primordial magnetic field with a strength of up to  $10^{-11}$~G.  Combination of the cosmic microwave background and gamma-ray constraints can thus sense the full range of possible cosmological magnetic fields to confirm or rule out their relevance to the problem of the origin of cosmic magnetic fields, as well as their influence on recombination and reionization epochs.  
\end{abstract}
%\maketitle

%%%%%%%%%%%%%%%%%%%%%%%%%%%%%
\section{Introduction}
%%%%%%%%%%%%%%%%%%%%%%%%%%%%%

\begin{figure}
\includegraphics[width=\linewidth]{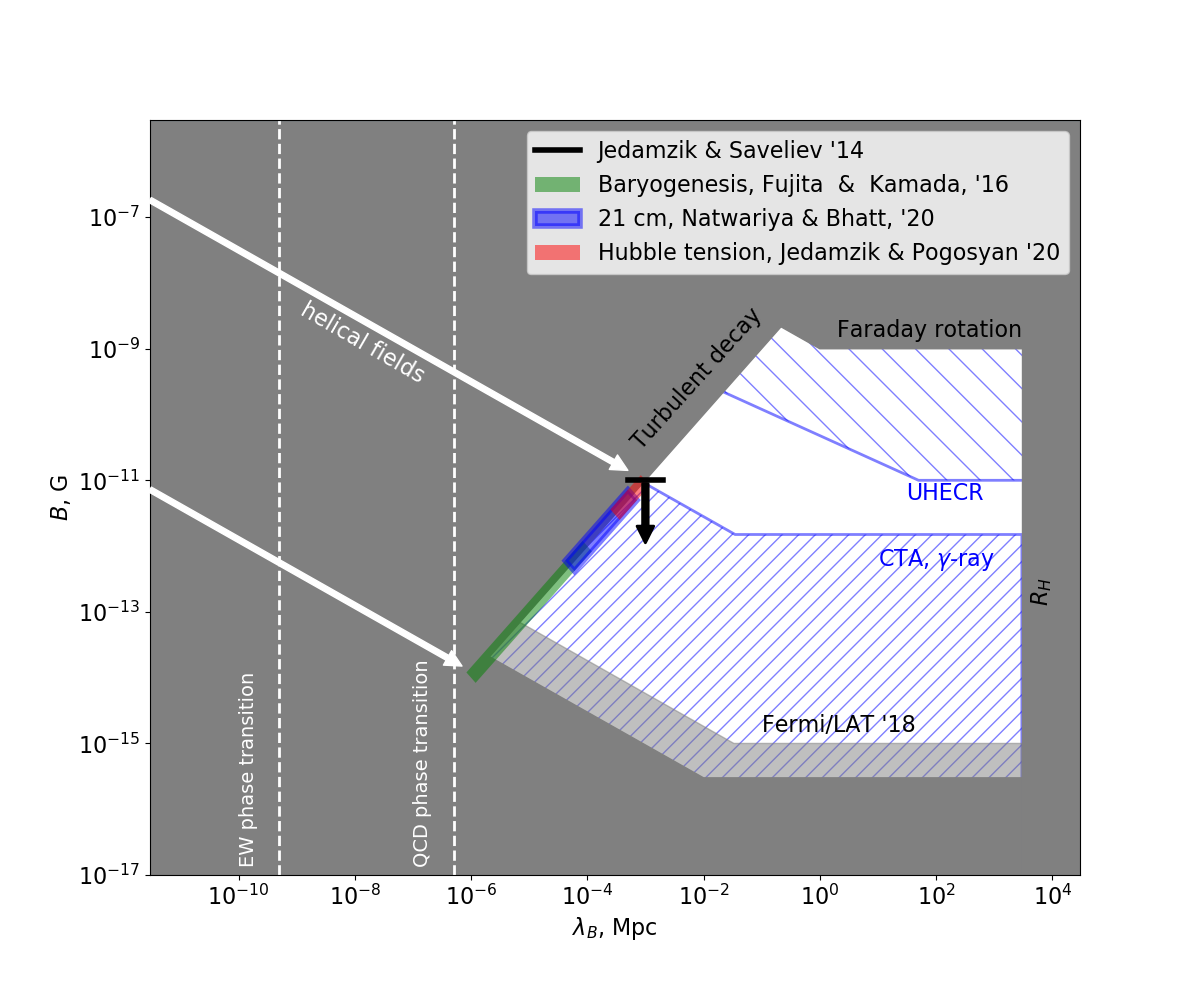}
\caption{Known constraints on the strength and correlation length of the IGMF \citep{neronov_semikoz09,durrer_neronov}. Red, blue, and green lines show the hints of the existence of a cosmological magnetic field from the CMB \citep{jedamzik_pogosyan}, 21 cm line \citep{edges1} and baryogenesis \citep{fujita16} correspondingly. Blue dashed regions show the sensitivity of different detection techniques \citep{neronov_semikoz09,durrer_neronov}. Black upper bound is from the analysis of the CMB signal by \citep{jedamzik_saveliev}.}
\label{fig:hints}
\end{figure}

Magnetic fields are present in almost every observable astronomical object. Yet their existence and role in the early universe is uncertain. A combination of \gr\ lower bounds \citep{neronov_vovk,fermi_bounds} and radio and cosmic microwave background (CMB) upper bounds \citep{kronberg94,planck16} on the intergalactic magnetic field (IGMF) provides evidence for the existence of fields with the strengths of $10^{-16}$~G$<B<10^{-9}$~G in the intergalactic medium (see Fig. \ref{fig:hints} for a summary of known constraints on IGMF, summarized by \citep{neronov_semikoz09,durrer_neronov}). However, the primordial nature of these fields has yet to be established. 

\citep{jedamzik_pogosyan} have recently shown that accounting for the magnetic field driven turbulence on plasma at the epoch of recombination modifies the estimate of the Hubble parameter from the CMB data and relaxes the $4.4\sigma$ tension between Planck measurements $H_0=67.36 \pm 0.54$~km/(s Mpc) based on $z\simeq 10^3$ data \citep{planck_H0} and present day universe measurements $H_0=74.03 \pm 1.42$~km/(s Mpc) using supernovae Type Ia \citep{sn_H0} and $H_0=73.3^{+1.7}_{-1.8}$`km/(s Mpc) based on gravitationally lensed systems \citep{lensing_H0}. The reasoning of \citep{jedamzik_pogosyan} futher develops the argument of \citep{jedamzik_saveliev} that a magnetic field present at the epoch of recombination induces clumping of baryonic matter and in this way modifies the recombination process. \citep{jedamzik_saveliev} have used this argument to derive a strong upper bound $B\lesssim 10^{-11}$~G at the epoch of recombination. This bound is shown by the black arrow in Fig. \ref{fig:hints}. The cosmological magnetic field with the strength $\sim 10^{-11}$~G is expected to have a correlation length of about $\lambda_B\simeq 1[B/10^{-11}\mbox{ G}]$~kpc, which is the largest processed eddy size at the epoch of recombination \citep{banerjee_jedamzik}.

%%%%%%%%%%%%%%%%%%%%%%%%%%%%%%%%%%%%%%%%%%%%%%%%
%%%%%%%%%%%%%%%%%%%%%%%%%%%%%%%%%%%%%%%%%%%%%%%%

Another indication of the existence of cosmological magnetic fields can be derived from measurements of redshift dependent absorption by the 21 cm line of neutral hydrogen at redshifts of about $z\sim 10$. The EDGES experiment has recently reported an excess opacity of the universe in the redshift range $15<z<20$ \citep{edges2}. This indicates that an efficient cooling mechanism of baryonic matter is operating during this epoch. Interaction of matter with magnetic field can provide both cooling (via pumping of thermal energy into magnetic field) and heating (via decay of MHD turbulence) of baryonic matter. It is possible to interpret the EDGES observation in terms of cooling through the interaction with magnetic fields \citep{edges1}. The magnetic field strengths, which is required to explain EDGES data, is comparable to that needed to resolve the Hubble parameter measurement tension, $5\times 10^{-13}$~G$<B<6\times 10^{-12}$~G \citep{edges1}. 

The magnetic fields surviving until the epochs of recombination and reionization should have been produced during phase transitions in the early universe (see \citep{durrer_neronov} for a review). The presence of a helical magnetic field at the epoch of electroweak phase transition can enable an explanation of the baryon asymmetry of the universe within the standard model of particle physics \citep{giovaninni97,fujita16}. The range of magnetic field strength $10^{-14}$~G$<B<10^{-12}$~G, which is compatible with this baryogenesis scenario is shown by the green shading in Fig. \ref{fig:hints}. Remarkably, the field strength required for a successful explanation of the baryon asymmetry is consistent with that needed for an explanation of the EDGES signal and of the Hubble parameter measurement tensions. 

The combination of these observational hints for the existence of a cosmological magnetic field defines an order-of-magnitude wide "sweet spot" around $B\sim 10^{-12}$~G in which the field estimates from multiple effects intersect. The most convincing evidence for the existence of a field with such strength would be its direct detection in the intergalactic medium. In what follows we explore the possibility of the measurement of such a field with \gr\ telescopes. We demonstrate that even though the field is at the upper sensitivity end of the \gr\ technique, its detection should still be possible with a deep exposure of the nearest blazars with CTA.  

%%%%%%%%%%%%%%%%%%%%%%%%%%%%%%%%%%%%%%
\section{Analytical estimates}
%%%%%%%%%%%%%%%%%%%%%%%%%%%%%%%%%%%%%%

Fields with strengths in the range of $B\sim 10^{-12}$~G are at the upper end of the sensitivity reach of the \gr\ measurement method \citep{neronov_semikoz09}. They are strong enough to deflect trajectories of electrons with energies in the 10-100~TeV range. This implies that the highest energy \gr\ signal accessible to telescopes should be used for the signal measurements. In this situation it is not clear if the small angle deflection approximation previously used for the sensitivity estimates used by \citep{neronov_semikoz09} is valid. We reassess the analytical estimates in this high-energy / strong field regime below. 

The correlation length $\lambda_B$ of cosmological magnetic fields scales with the strength as \citep{banerjee_jedamzik}
\begin{equation}
\label{eq:lambda}
    B\sim 10^{-11}\left[\frac{\lambda_B}{1\mbox{ kpc}}\right]\mbox{ G}
\end{equation}
The field power spectrum is shaped by the turbulence on the scales shorter than $\lambda_B$. The integral length scale of the field with power spectrum $P_B(k)\propto k^{-n}$ as
\begin{equation}
\label{eq:L}
    \lambda_B \approx L_B \frac{n-1}{2n} = \frac{L_B}{5}
\end{equation}
for $n=5/3$ (assuming the Kolmogorov turbulence spectrum), where $L_B$ is the maximum scale of the Kolmogorov spectrum.

We consider secondary emission induced by interactions of primary \gr s with energies $E_{\gamma0}$. The mean free path of these \gr s through the EBL is 
\begin{equation}
    \lambda_{\gamma 0}\simeq 2.5\left[\frac{E_{\gamma 0}}{100\mbox{ TeV}}\right]^{-1.6}\mbox{ Mpc}
\end{equation}
For the analytical estimates we assume that each primary \gr\ produces an electron and a positron with energies of $E_e=E_{\gamma 0}/2$. The electrons and positrons cool due to the inverse Compton scattering of CMB photons on the distance scale
\begin{equation}
    D_e\simeq 7
    \left[\frac{E_e}{50\mbox{ TeV}}\right]^{-1}\mbox{ kpc}
\end{equation}
in the Thomson regime of inverse Compton scattering\footnote{Our numerical modeling takes into account the full Klein-Nishina  cross-section of inverse Compton scattering. The Klein-Nishna effect corrects analytical estimates, but does not change the qualitative picture presented in this section.}  relevant for the scattering of CMB photons by electrons with energies $\lesssim 100$~TeV. Such electrons produce inverse Compton emission at the energy 
\begin{equation}
    E_\gamma\simeq 8\left[\frac{E_e}{50\mbox{ TeV}}\right]^2\mbox{ TeV}
\end{equation}
The gyroradius of electrons is
\begin{equation}
    R_L\simeq 5\left[\frac{E_e}{50\mbox{ TeV}}\right]\left[\frac{B}{10^{-11}\mbox{ G}}\right]^{-1}\mbox{ kpc}
\end{equation}
Electrons are typically deflected by an angle of
\begin{eqnarray}
    &&\delta=\frac{\lambda_B}{R_L}\simeq 0.06\left[\frac{E_e}{50\mbox{ TeV}}\right]^{-1}\left[\frac{B}{10^{-11}\mbox{ G}}\right]^2
\end{eqnarray}
on the distance scale about the integral length of the magnetic field (we have used the scaling $B\propto \lambda_B$ suggested by Eqs. (\ref{eq:lambda}) and (\ref{eq:L}).  Accumulation of such small deflections on the electron cooling distance scale $D_e$ results in the overall deflection
\begin{eqnarray}
    &&\Delta=\sqrt{\frac{D_e}{\lambda_B}}\delta\simeq 
    0.2\left[\frac{E_e}{50\mbox{ TeV}}\right]^{-3/2}\left[\frac{B}{10^{-11}\mbox{ G}}\right]^{3/2}\nonumber\\
    &&\simeq 0.2\left[\frac{E_\gamma}{8\mbox{ TeV}}\right]^{-3/4}\left[\frac{B}{10^{-11}\mbox{ G}}\right]^{3/2}
\end{eqnarray}

If the field strength is $B\sim 10^{-11}$~G, the opening angle of the secondary emission cone at 2~TeV can be as large as the opening angle of the AGN jets. The secondary emission flux within the cone gets suppressed as 
\begin{eqnarray}
    &&\frac{F_{ext}}{F_{\gamma 0}}=\frac{\Theta_{jet}^2}{\Delta^2}\simeq1\left[\frac{E_\gamma}{8\mbox{ TeV}}\right]^{3/2}\\
    &&\left[\frac{B}{10^{-11}\mbox{ G}}\right]^{-3}\left[\frac{\Theta_{jet}}{10^\circ}\right]^2, \ \ \ B\gtrsim 10^{-11}\mbox{ G}\nonumber
    \label{eq:fext}
\end{eqnarray}
where we have assumed $\Theta_{jet}\sim 10^\circ\simeq 0.2$. Such flux suppression occurs below the energy  at which $\Delta=\Theta_{jet}$,
\begin{equation}
\label{eq:ecrit}
    E_{crit}=8\left[\frac{B}{10^{-11}\mbox{ G}}\right]^2\left[\frac{\Theta_{jet}}{10^\circ}\right]^{-4/3}\mbox{ TeV}
\end{equation}

If $B< 10^{-11}$~G, the deflection angle $\Delta$ is smaller than the opening angle of the jet and extended emission is still observable toward 10 TeV energy. 

The maximal possible angular size of the extended emission is determined by the transverse size of the jet at the distance $\lambda_{\gamma0}$
\begin{equation}
    R_{ext,max}=\Theta_{jet}\lambda_{\gamma0}\simeq 0.5\left[\frac{\Theta_{jet}}{10^\circ}\right]\left[\frac{E_{\gamma0}}{100\mbox{ TeV}}\right]^{-1.6}\mbox{ Mpc}
\end{equation}
This corresponds to the angular size
\begin{eqnarray}
    &&\Theta_{ext,max}=\frac{R_{ext}}{D}\simeq 0.24^\circ \left[\frac{\Theta_{jet}}{10^\circ}\right]\left[\frac{E_{\gamma0}}{100\mbox{ TeV}}\right]^{-1.6}\\
    &&\left[\frac{D}{120\mbox{ Mpc}}\right]^{-1}\simeq 0.24^\circ \left[\frac{\Theta_{jet}}{10^\circ}\right]\left[\frac{E_{\gamma}}{8\mbox{ TeV}}\right]^{-0.8}\left[\frac{D}{120\mbox{ Mpc}}\right]^{-1} \nonumber
\end{eqnarray}
where we have used the distances to Mrk 421 and Mrk 501 for the numerical estimate. The time delay of the extended signal can be estimated as 
\begin{eqnarray}
    &&T_{ext,max}=D\Theta_{ext,max}^2/c\simeq\\ &&8\left[\frac{\Theta_{jet}}{10^\circ}\right]^2\left[\frac{E_{\gamma}}{8\mbox{ TeV}}\right]^{-1.6} \mbox{ kyr}\nonumber
\end{eqnarray}
This time scale imposes a requirement on the duty cycle of the source for which the extended emission is detectable: the source should have been active over the last 10~kyr, which is plausible given the fact that the source posesses a kiloparsec scale jet \citep{kpc_jet}. 

Note that the maximal extended source size does not depend on the magnetic field strength. Instead, it is the extended source flux that scales with magnetic field. However, this dependence could hardly be used for the estimation of the field strength because of uncertainty of the opening angle of the jet and because of the very rapid decline of the extended source flux with the field strength. Because of this limitation, we choose to show in Fig. \ref{fig:hints} the lower bound for the Fermi/LAT telescope derived in the regime of small deflection angles of electrons (see \citep{fermi_bounds} for details). 

%%%%%%%%%%%%%%%%%%%%%%%%%%%%%%%%%%%%%%
\begin{figure}
    \includegraphics[width=\linewidth]{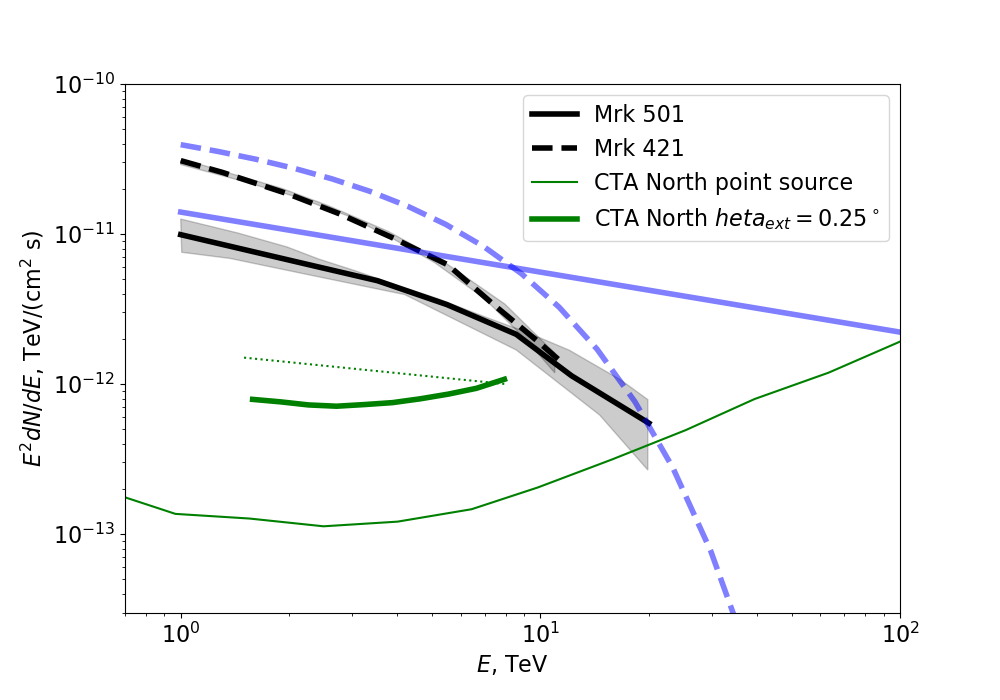}
    \caption{Comparison of high-energy ends of the spectra of Mrk 421 and Mrk 501 \citep{mrk_hawc}. Black curves are observed, and blue curves are the intrinsic spectra of the sources. Thin and thick green lines show CTA North sensitivity to point sources and to the extended sources with angular size $\theta = 0.25^\circ$ \footnote{https://www.cta-observatory.org/science/cta-performance/}.  Green dotted line shows an analytical estimate of the secondary \gr\ flux from Mrk 501 in 1-10 energy range assuming no influence of IGMF.  }
    \label{fig:mrk501}
\end{figure}
%%%%%%%%%%%%%%%%%%%%%%%%%%%%%%%%%%%%%%

%%%%%%%%%%%%%%%%%%%%%%%%%%%%%%%%%%%%%%
\section{Selection of best target for the search of strong IGMF}
%%%%%%%%%%%%%%%%%%%%%%%%%%%%%%%%%%%%%%

Probe of the strongest fields $B\lesssim 10^{-11}$~G requires 
\begin{enumerate}
    \item[(a)] large primary point-source power in the 100 TeV energy range,
    \item[(b)] detectability of extended emission in multi-TeV energy range, and
    \item[(c)] presence of primordial IGMF in the several Mpc region around the source.
\end{enumerate}

Below we present arguments that at least one source, Mrk 501, fulfills these three conditions and can be used for the probe of strong IGMF of cosmological origin.

%%%%%%%%%%%%%%%%%%%%%%%%%%%%%%%%%%%%%%
\subsection{Are there 100 TeV bright blazars?}
%%%%%%%%%%%%%%%%%%%%%%%%%%%%%%%%%%%%%%

There are currently no measurements or estimates of the blazar luminosities at 100 TeV. The highest energy photons detected from blazars are those from the two nearest BL Lac objects, Mrk 421 and Mrk 501. Fig. \ref{fig:mrk501}  shows the spectral energy distribution of these sources measured by HAWC \citep{mrk_hawc}. One can see that even though Mrk 421 is the brightest source at TeV, it has a softer spectrum and its intrinsic luminosity is most probably strongly suppressed at 100 TeV. To the contrary, Mrk 501 has a harder spectrum that does not show any signature of high-energy cutoff. The spectrum is measured up to 20 TeV. In view of this fact, we consider Mrk 501 as a more promising candidate for the search of the strongest IGMF and the following calculations are limited to this source. We assume that its intrinsic spectrum extends up to 100 TeV, as shown in Fig. \ref{fig:mrk501}. 

Extrapolating the power law measured by HAWC up to 100~TeV, we find that the intrinsic flux of the source at this energy should be at the level of 
\begin{equation}
    F_{\gamma 0}(100 \mbox{ TeV})\simeq (2.5\pm 1)\times 10^{-12}\mbox{ TeV/(cm}^2\mbox{s)}
\end{equation}

%%%%%%%%%%%%%%%%%%%%%%%%%%%%%%%%%%%%%%
\subsection{Detectability of multi-TeV extended emission}
%%%%%%%%%%%%%%%%%%%%%%%%%%%%%%%%%%%%%%

Most of the intrinsic flux in the 100 TeV energy range is absorbed in the intergalactic medium and is converted to the extended emission. In the regime $B<10^{-11}$~G, the extended emission flux is not suppressed by the widening of the opening angle of the emission cone and one can expect that the flux of the extended source is 
\begin{equation}
    F_{ext}(8 \mbox{ TeV})\sim F_{\gamma 0}(100 \mbox{ TeV})/2\sim 10^{-12}\mbox{ TeV/(cm}^2\mbox{s)}
\end{equation}
This flux level is well above the CTA point-source sensitivity, shown by the lowermost green solid curve in Fig. \ref{fig:mrk501}. The sensitivity is limit $F_{CTA,ps}\simeq 5\times 10^{-14}$~TeV/(cm$^2$s) in the 1-10 TeV range is determined by the statistical fluctuations of the background within the point spread function $\Theta_{psf}\simeq 0.04^\circ$\footnote{https://www.cta-observatory.org/science/cta-performance/}. The sensitivity for extended source flux worsens with the increase of the source size, roughly as 
\begin{equation}
    F_{CTA,ext}=F_{CTA,ps}\frac{\Theta_{ext}}{\Theta_{PSF}}\simeq 3\times 10^{-13}\left[\frac{\Theta_{ext}}{0.25^\circ}\right]\mbox{ TeV/(cm}^2\mbox{s)}
\end{equation}
This sensitivity is compared with the expected extended source flux in Fig. \ref{fig:mrk501}. As one can see, the extended flux is above the 
sensitivity limit and might be detectable. However, the 
extended emission appears on top of the point-source flux and more detailed analysis is required for verification of its detectability.

%%%%%%%%%%%%%%%%%%%%%%%%%%%%%%%%%%%%%%
\subsection{IGMF in the direction of Mrk 501}
%%%%%%%%%%%%%%%%%%%%%%%%%%%%%%%%%%%%%%

The \gr\ measurement method is sensitive to the weakest volume filling IGMF along the line of sight toward the source. This field is typically found in the voids of the LSS. TeV \gr s can cross many voids before being absorbed in pair production. This is not the case for the 100~TeV \gr s, which can potentially be used to probe the strongest IGMF $B\sim 10^{-11}$~G. These \gr s can only probe the field within several megaparsecs around the source. Indeed, in this case the effect of extended emission is most significant at energies of secondary photons in the range of $\sim$ 2 - 10 TeV. The corresponding energy of primary photons is about 50 - 100 TeV. Considering the standard EBL models the typical mean free path of such photons is 3 - 10 Mpc. If the source is occasionally found in or near a large galaxy cluster, or the line of sight points toward a filament of the LSS, the field strength can be much higher than the typical void field making the line of reasoning described above inapplicable.

To explore environmental effects around Mrk 501, we rely on constrained cosmological simulations derived using the Bayesian Origin Reconstruction from Galaxies (BORG) inference method \citep{borg_jw_13}. BORG provides a statistical ensemble of initial conditions that all match detailed observations of galaxy count in the universe. To achieve this, it relies on modeling the evolution of large-scale structures starting from high redshift using a dynamical model. Then the probability of such a sample is compared volume element by volume element to data from a galaxy survey assuming some bias function and Poisson statistics for the number count \citep{borg_jw_13}. The initial conditions can then be resimulated with a software of the user's choice. For the data considered here, the dynamical model is a particle mesh $N$-body solver with 20 timesteps; the data are provided by the 2M++ galaxy compilation \citep{Lavaux2011_TMPP}. The resolution of the density contrast of the initial conditions and at $z=0$ are set to $2.64h^{-1}$~ Mpc. The details of the algorithm and of the inference procedure are provided in \citep{Jasche2019}. 

%%%%%%%%%%%%%%%%%%%%%%%%%%%%%%%%%%%%%%
\begin{figure}
    \includegraphics[width=\linewidth]{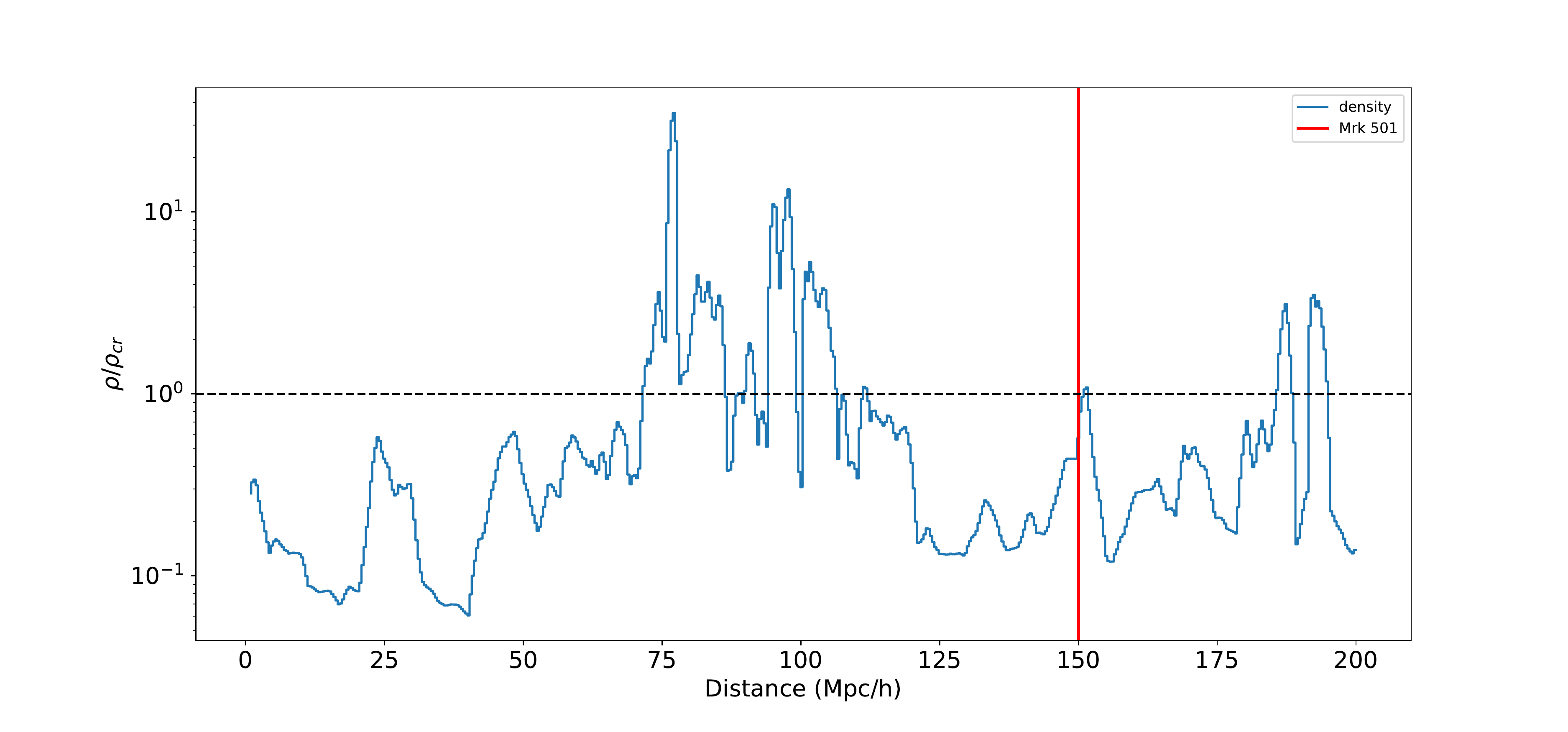}
    \caption{Dark matter density profile along the line of sight toward Mrk 501 based on constrained simulations of the LSS \citep{Jasche2019}.}
    \label{fig:mrk501_map}
\end{figure}
%%%%%%%%%%%%%%%%%%%%%%%%%%%%%%%%%%%%%%

%%%%%%%%%%%%%%%%%%%%%%%%%%%%%%%%%%%%%%
\begin{figure*}
    \includegraphics[width=\textwidth]{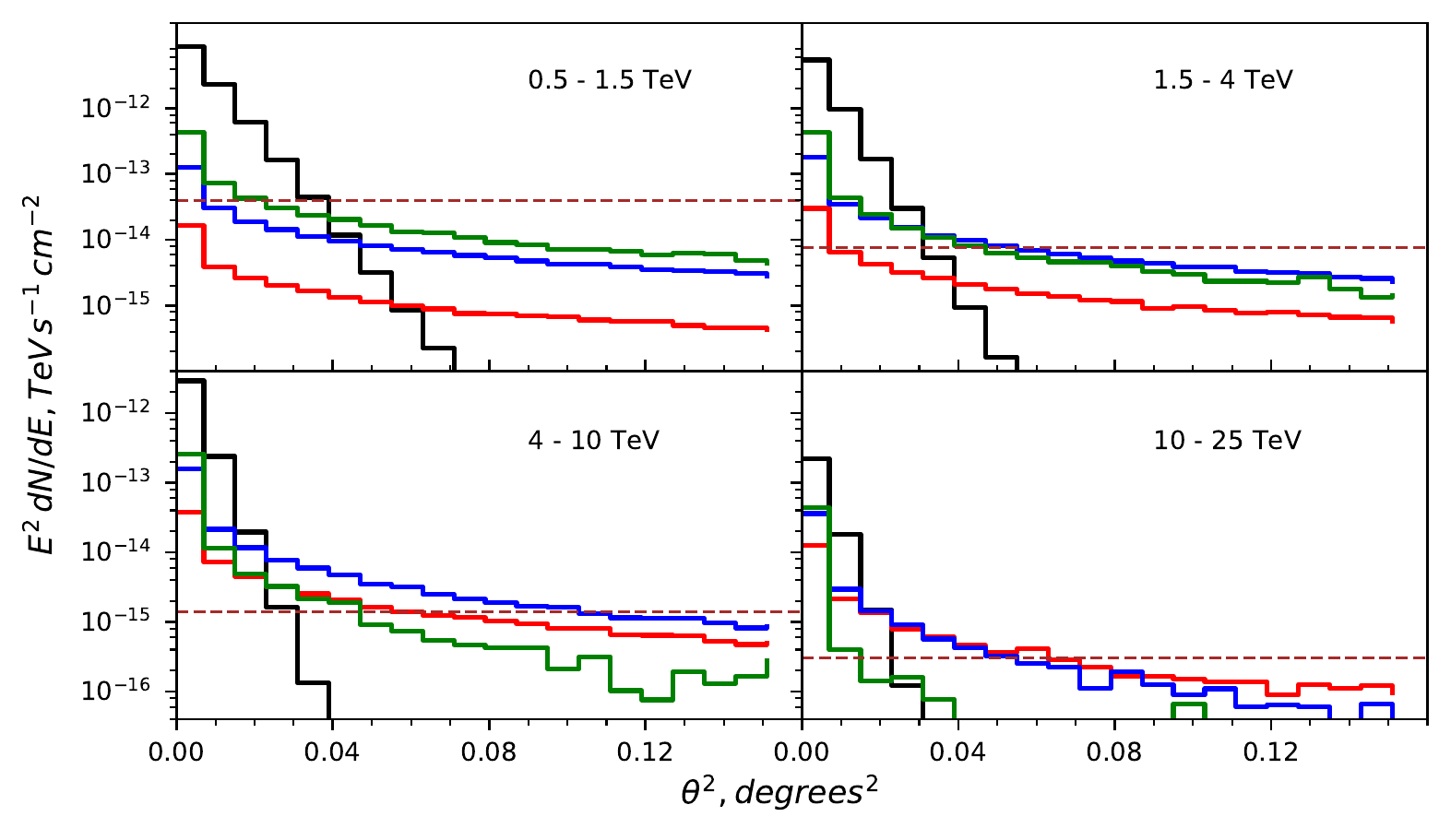}
    \caption{Angular distribution of primary and secondary photons in different energy ranges. Black histograms show the primary point-source signal; green, blue, and red histograms show the extended emission calculated for different magnetic field strengths: $10^{-12}$~G, $3\times 10^{-12}$~G, and $10^{-11}$~G. Horizontal dashed line shows the level of residual cosmic-ray electron background measured by the HESS \citep{hess_icrc2017}.}
    \label{fig:mrk501_ang_spec}
\end{figure*}
%%%%%%%%%%%%%%%%%%%%%%%%%%%%%%%%%%%%%%

Using the dark matter density as an estimator of the strength of magnetic field we found (see Fig. \ref{fig:mrk501_map}) that the IGMF around the source does not appear in the reach galaxy cluster. Instead there is a large void just next to the Mrk 501 with the average density $\rho$ below the critical density of the universe and there the magnetic field is not amplified in the course of structure formation. Numerical modeling of the IllustrisTNG simulation \citep{illistristng} shows that such amplification only occurs in the regions with overdensity $\rho/\rho_{cr}\gtrsim 10$.

%%%%%%%%%%%%%%%%%%%%%%%%%%%%%%%%%%%%%%
\section{Numerical Modeling}
%%%%%%%%%%%%%%%%%%%%%%%%%%%%%%%%%%%%%%

Qualitative arguments presented in the previous section show that even if the IGMF of cosmological origin  is as strong as $B\sim 10^{-11}$~G, it should still be possible to detect it via the search of extended emission around Mrk 501, which is the brightest blazar in the 10 TeV sky. 

In this section we support this qualitative argument with numerical modeling of the extended source signal. For this purpose we use the Monte Carlo simulation code developed in~\citep{Berezinsky:2016feh} which was also tested by comparison with the alternative cascade simulations~\citep{Taylor:2011bn,Kalashev:2014xna,Kachelriess:2011bi}.

We consider a primary \gr\ source with the power-law spectrum with the slope $dN/dE\propto E^{-2.4}$ extending up to 100~TeV energy, situated at the distance of $D=150$~Mpc. The \gr s are emitted into a jet with opening angle $\Theta_{jet}=5^\circ$ aligned along the line of sight. 

We propagate the primary \gr\ beam toward the observer, taking into account pair production on extragalactic background light (EBL). The EBL spectrum is that of \citet{Gilmore:2011ks}. Electrons and positions produced by absorbed \gr s loose energy due to inverse Compton scattering on EBL and CMB. The IGMF present that deflects electrons and positrons is assumed to have strength and correlation length satisfying the relation (\ref{eq:lambda}). 

The field is generated with the Kolmogorov power spectrum, following the method of \citet{jokipii99} and constructed as a sum of finite number of randomly oriented plane waves. Wavenumbers were evenly distributed in logarithmic scale between $k_{min}$ and $k_{max}$. The number of modes was set to 500 and ($k_{max}/k_{min})=100$. The exact value of $k_{max}$ depends on magnetic field strength and is determined from (\ref{eq:lambda}) and (\ref{eq:L}) to properly fit the correlation length. We have checked that in the small scattering regime we obtain the theoretically expected result from the random walk process with a correlation length equal to $\lambda_B$.

Fig. \ref{fig:mrk501_ang_spec} shows the result of the calculation of the extended emission pattern at different energies for a range of IGMF strengths. The extended emission signal appears on top of the point-source signal and of the residual charged cosmic-ray background. The difficulties of detection of the extended emission are evident from the figure. The extended source signal is always subdominant. The point-source signal dominates in the $0^\circ<\theta<0.17^\circ$ angular range, while the residual cosmic-ray background dominates outside $\theta_c = \sqrt{0.03}\simeq 0.17^\circ$.

%%%%%%%%%%%%%%%%%%%%%%%%%%%%%%%%%%%%%%
\begin{figure*}
    \includegraphics[width=\textwidth]{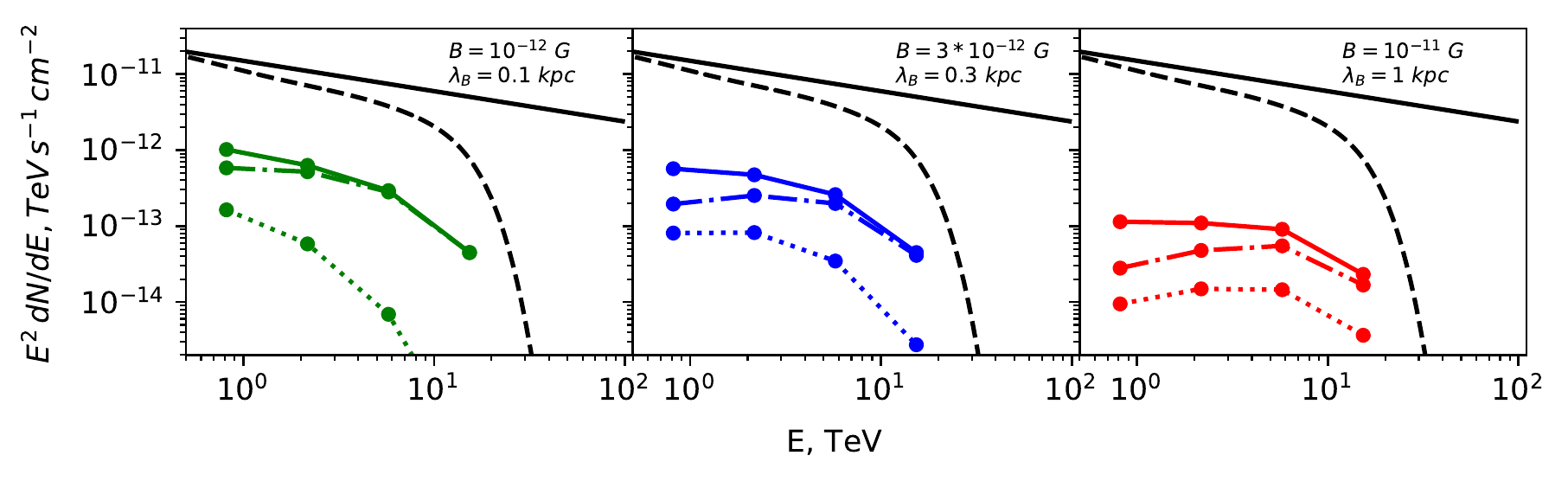}
    \caption{Intrinsic spectrum of Mrk 501 (black solid line) in comparison with the absorbed spectrum (black dashed line) and the spectrum of secondary emission (colored lines) for different parameters of the magnetic field. Solid colored line corresponds to the total flux of the secondary emission from the source, while dashed-dotted and dotted lines are the fluxes in the angular ranges [$0:\theta_c$] and [$\theta_c:0.4$] degrees. }
    \label{fig:mrk501_E_spec}
\end{figure*}
%%%%%%%%%%%%%%%%%%%%%%%%%%%%%%%%%%%%%%

Fig. \ref{fig:mrk501_E_spec}, which is a direct analog of the analytical estimate in figure \ref{fig:mrk501},  provides a further illustration of the dominance of the point-source flux. The extended source spectrum is split into two components, outside and inside $\theta_c$. In the $B\sim 10^{-12}$~G case, the extended source flux is largely dominated by the component inside $\theta_c$ (and hence, within the extent of the primary source point spread function). If $B\sim 10^{-11}$~G, emission flux outside $\theta_c$ is just a factor of 2 below the flux within $\theta_c$. Given the overall weakness of the extended emission signal in the case of a strong field, it is thus reasonable to use the combination of regions $\theta<\theta_c$ and $\theta>\theta_c$, searching for the deviations from the PSF of the instrument in the angular range $[0^\circ: 0.4^\circ]$. 

%%%%%%%%%%%%%%%%%%%%%%%%%%%%%%%%%%%%%%
\begin{figure}
    \includegraphics[width=\linewidth]{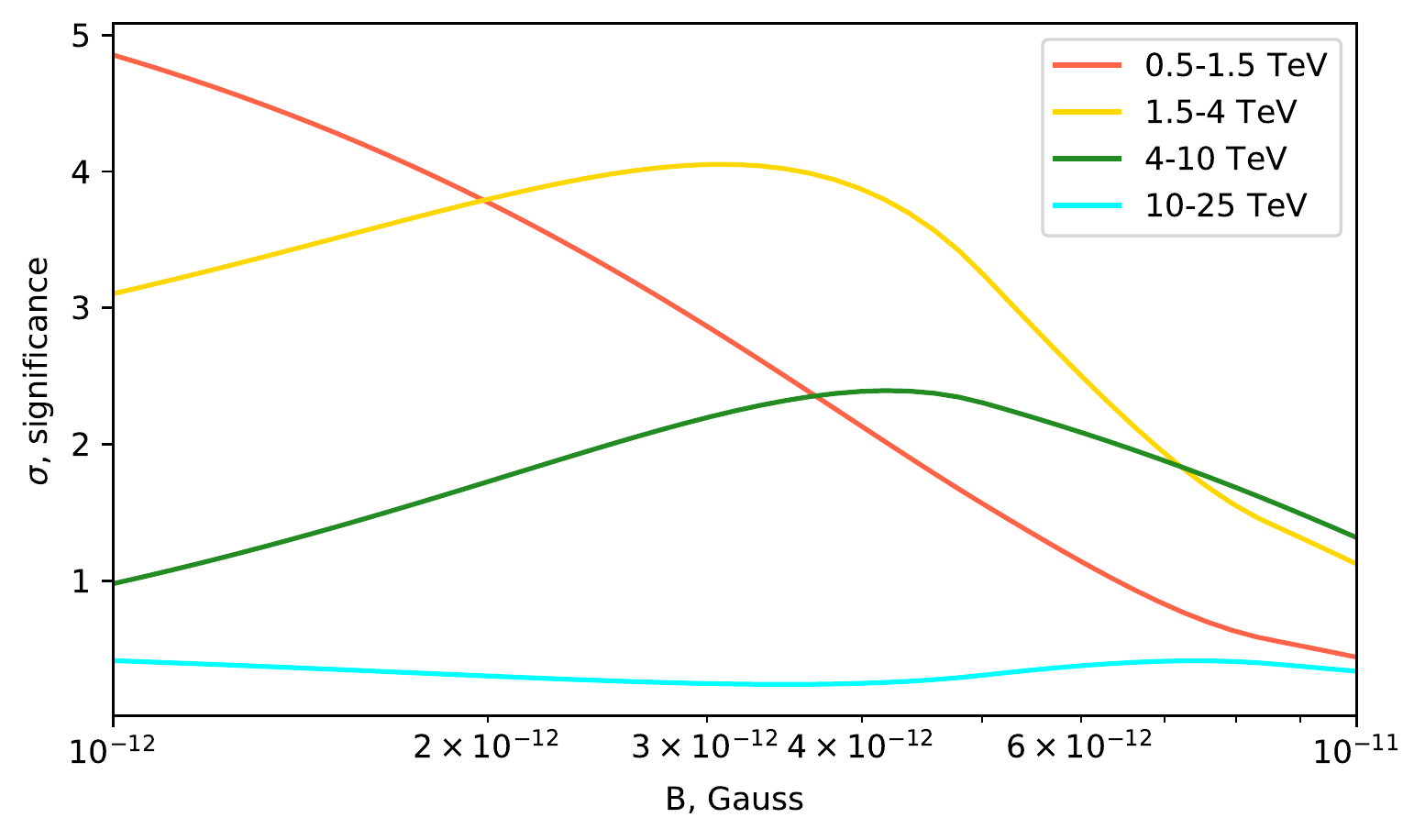}
    \caption{Significance of detection of the extended emission signal in different energy ranges as a function of the assumed magnetic field strength. The assumed exposure of CTA is $T=50$~hr.}
    \label{fig:mrk501_bins_signif}
\end{figure}
%%%%%%%%%%%%%%%%%%%%%%%%%%%%%%%%%%%%%%

%%%%%%%%%%%%%%%%%%%%%%%%%%%%%%%%%%%%%%
\begin{figure}
    \includegraphics[width=\linewidth]{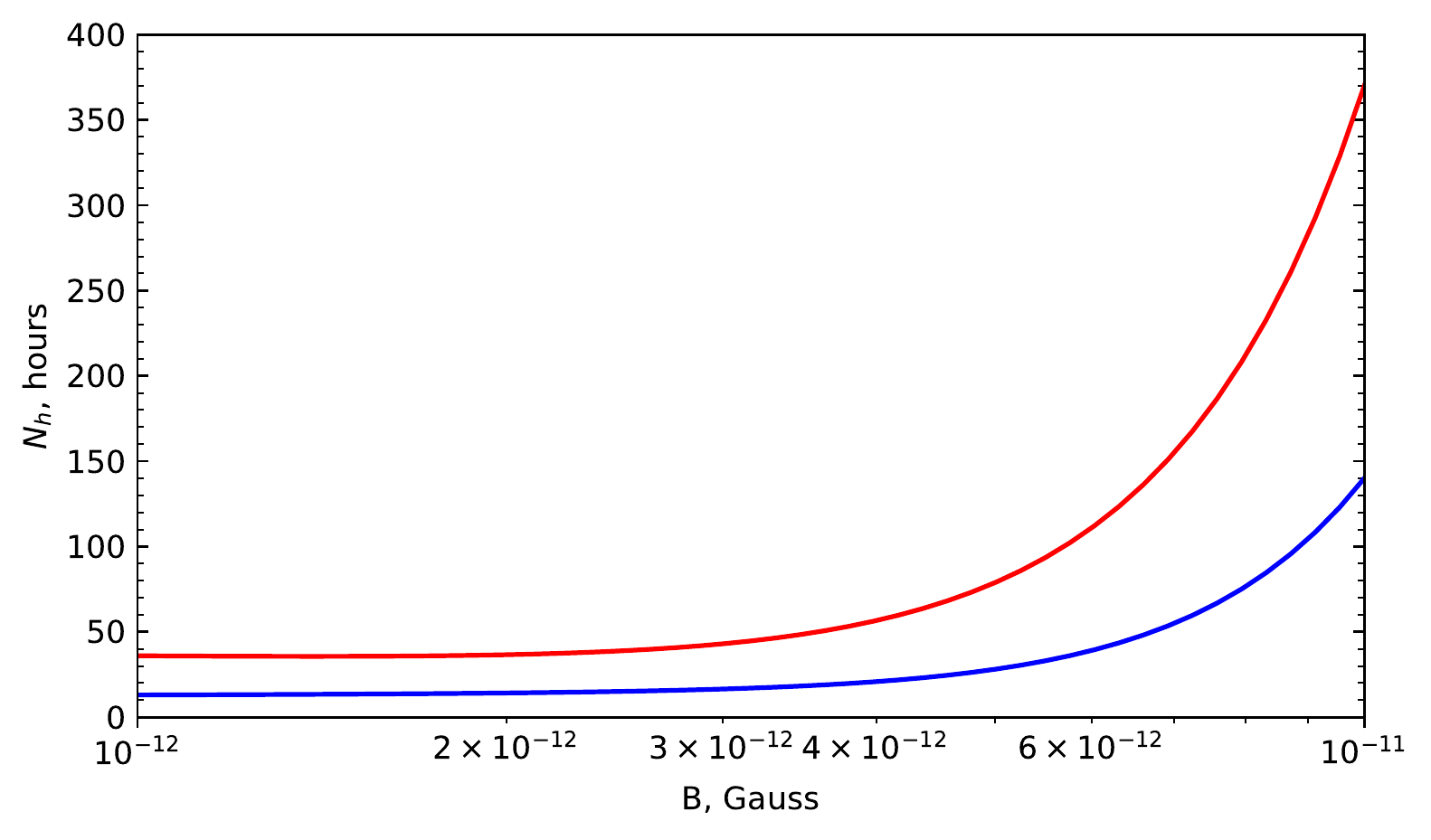}
    \caption{CTA exposure time needed for $3\sigma$ (lower curve) and $5\sigma$ (upper curve) detection of the extended emission signal, as a function of IGMF strength.}
    \label{fig:mrk501_hours}
\end{figure}
%%%%%%%%%%%%%%%%%%%%%%%%%%%%%%%%%%%%%%

Fig. \ref{fig:mrk501_E_spec} also illustrates the effect of the overall suppression of the extended source flux below $E_{crit}$ (\ref{eq:ecrit}) in the strong IGMF case. This is most clearly seen in the right panel of the figure. In this case $E_{crit}=8$~TeV for the assumed jet opening angle. Comparing the extended source flux above 10~TeV for the $B=10^{-11}$~G case with that for the $B=10^{-12}$~G case (shown in the left panel) cone can see that the total extended source fluxes are comparable above 10~TeV. This is not the case for the 1~TeV band fluxes. The opening angle of the secondary emission cone at 1 TeV in the case of the $B\sim 10^{-11}$~G field is wider than the jet opening angle and the flux is suppressed by a factor of 10. 

%%%%%%%%%%%%%%%%%%%%%%%%%%%%%%%%%%%%%%
\section{Results and discussion}
%%%%%%%%%%%%%%%%%%%%%%%%%%%%%%%%%%%%%%

We have used the results of the Monte Carlo modelling to investigate the detectability of the extended emission signal with CTA. To do this, we have calculated the statistics of the point-source signal,  extended emission signal, and residual cosmic-ray background in each angular bin of the histograms shown in Fig. \ref{fig:mrk501_ang_spec}, for different CTA exposures. In this way we have generated mock CTA datasets. We have fitted the mock data with a model of point source plus the residual cosmic-ray background model, ignoring the presence of the extended source. We have then estimated the significance of detection of the extended emission in the simulated data set by calculating the level of inconsistency of the "point source + residual cosmic ray electron background \citep{hess_icrc2017}" model with the simulated data. 

The results of this analysis are shown in Fig. \ref{fig:mrk501_bins_signif}. If the IGMF strength is below $3\times 10^{-12}$~G, the extended signal is detectable with a significance larger than $3\sigma$ in the energy ranges 0.5-1.5 TeV and $1.5-4$~TeV. A somewhat stronger magnetic field, $3\times 10^{-12}<B<6\times 10^{-12}$~G is still marginally detectable through the extended emission at somewhat higher energy, up to 10~TeV. The extended emission signal associated with the $10^{-11}$~G IGMF is not detectable in a 50~hr exposure. 

Fig. \ref{fig:mrk501_hours} shows the exposure needed for the $3\sigma$ evidence for and $5\sigma$ discovery of the extended emission for different IGMF strengths. From this figure one can see that with a 150 hr exposure, an evidence for the presence of extended emission in the 10~TeV energy range can be found even for the magnetic field with the strength of $10^{-11}$~G. Extended emission shaped by deflections of electrons and positrons in such IGMF can be discovered in such a very long exposure, $T\simeq 400$~hr. 

CTA is a unique instrument that will be capable of detecting the extended emission in the multi-TeV energy range. Next generation air shower arrays, the Large High Altitude Air Shower Observatory (LHAASO) \citep{lhaaso} and the Southern Wide-field Gamma-ray Observatory (SWGO) \citep{swgo}, Tunka Advanced Instrument for cosmic rays and Gamma Astronomy (TAIGA) \citep{taiga} do not achieve a comparable level of suppression of charged cosmic-ray background in the energy range below 10~TeV and have much worse angular resolution in this energy range. Both factors strongly reduce sensitivity for the extended emission search.

Mrk 501 is the brightest extragalactic source in the 10 TeV energy range. Its uniqueness can justify such an ultra-deep exposure (perhaps accumulated over several years of observations). It can serve not only for the measurement of the IGMF but also for the study of the high-energy end of blazar spectra and for the precision measurements of the EBL.  

Optimization of CTA performance for the search of extended emission around Mrk 501 will build upon previous developments with the current generation Cerenkov telescopes. Recent measurements of extension of the Crab nebula show the potential of such a search \citep{crab_ang_size}. Mrk 421 and Mrk 501, two sources with TeV band fluxes comparable to Crab have already been subjects of exteneded emission searches with MAGIC. \cite{magic} have derived constraints on the extended emission flux at the level of 5\% of the Crab nebula. HESS has derived a tight constraint of 30$\arcsec$ on the angular size of the source generating the bulk of Mrk 421 flux. The source emission in this angular range is dominated by primary photons rather than by the cascade emission.

To summarize, we have shown that direct detection of the strong cosmological magnetic field, which is needed for the resolution of the tension between different measurements of the Hubble parameter \citep{jedamzik_pogosyan} and for the explanation of the EDGES 21 cm line opacity \citep{edges1} is possible with the \gr\ measurement technique. At least one source, Mrk 501, is suitable for this purpose. A search for the extended emission around Mrk 501 in the 1-10 TeV energy range can result in the detection of IGMF with a the strength of up to $10^{-11}$~G and a correlation length of up to 10~kpc. 

\section*{Acknowledgements}

We thank Guilhem Lavaux and Jens Jasche for providing the constrained simulations of the Local Universe derived using the BORG algorithm.  We also thank Karsten Jedamzik and Levon Pogosyan for clarifying the details of the estimate of cosmological magnetic field from their analysis of the CMB data. 

This work has been supported by the French National Research Agency (ANR) grant ANR-19-CE31-0020 and Russian Science Foundation grant 20-42-09010.

\bibliography{references}{}

\begin{thebibliography}{}
\expandafter\ifx\csname natexlab\endcsname\relax\def\natexlab#1{#1}\fi
\providecommand{\url}[1]{\href{#1}{#1}}
\providecommand{\dodoi}[1]{doi:~\href{http://doi.org/#1}{\nolinkurl{#1}}}
\providecommand{\doeprint}[1]{\href{http://ascl.net/#1}{\nolinkurl{http://ascl.net/#1}}}
\providecommand{\doarXiv}[1]{\href{https://arxiv.org/abs/#1}{\nolinkurl{https://arxiv.org/abs/#1}}}

\bibitem[{{Ackermann} {et~al.}(2018)}]{fermi_bounds}
{Ackermann}, M., {et~al.} 2018, \apjs, 237, 32,
  \dodoi{10.3847/1538-4365/aacdf7}

\bibitem[{Aghanim {et~al.}(2020)}]{planck_H0}
Aghanim, N., {et~al.} 2020, Astron. Astrophys., 641, A6,
  \dodoi{10.1051/0004-6361/201833910}

\bibitem[{Albert {et~al.}(2019)}]{swgo}
Albert, A., {et~al.} 2019.
\newblock \doarXiv{1902.08429}

\bibitem[{{Aleksi{\'c}} {et~al.}(2010){Aleksi{\'c}}, {Antonelli}, {Antoranz},
  {Backes}, {Baixeras}, {Barrio}, {Bastieri}, {Becerra Gonz{\'a}lez},
  {Bednarek}, {Berdyugin}, {Berger}, {Bernardini}, {Biland}, {Blanch}, {Bock},
  {Bonnoli}, {Bordas}, {Borla Tridon}, {Bosch-Ramon}, {Bose}, {Braun}, {Bretz},
  {Britzger}, {Camara}, {Carmona}, {Carosi}, {Colin}, {Commichau}, {Contreras},
  {Cortina}, {Costado}, {Covino}, {Dazzi}, {de Angelis}, {de Cea Del Pozo}, {de
  Los Reyes}, {de Lotto}, {de Maria}, {de Sabata}, {Delgado Mendez}, {Doert},
  {Dom{\'\i}nguez}, {Dominis Prester}, {Dorner}, {Doro}, {Elsaesser},
  {Errando}, {Ferenc}, {Fonseca}, {Font}, {Garc{\'\i}a L{\'o}pez},
  {Garczarczyk}, {Gaug}, {Godinovic}, {Hadasch}, {Herrero}, {Hildebrand },
  {H{\"o}hne-M{\"o}nch}, {Hose}, {Hrupec}, {Hsu}, {Jogler}, {Klepser},
  {Kr{\"a}henb{\"u}hl}, {Kranich}, {La Barbera}, {Laille}, {Leonardo},
  {Lindfors}, {Lombardi}, {Longo}, {L{\'o}pez}, {Lorenz}, {Majumdar}, {Maneva},
  {Mankuzhiyil}, {Mannheim}, {Maraschi}, {Mariotti}, {Mart{\'\i}nez}, {Mazin},
  {Meucci}, {Miranda}, {Mirzoyan}, {Miyamoto}, {Mold{\'o}n}, {Moles},
  {Moralejo}, {Nieto}, {Nilsson}, {Ninkovic}, {Orito}, {Oya}, {Paiano},
  {Paoletti}, {Paredes}, {Partini}, {Pasanen}, {Pascoli}, {Pauss}, {Pegna},
  {Perez-Torres}, {Persic}, {Peruzzo}, {Prada}, {Prandini}, {Puchades},
  {Puljak}, {Reichardt}, {Rhode}, {Rib{\'o}}, {Rico}, {Rissi}, {R{\"u}gamer},
  {Saggion}, {Saito}, {Salvati}, {S{\'a}nchez-Conde}, {Satalecka}, {Scalzotto},
  {Scapin}, {Schultz}, {Schweizer}, {Shayduk}, {Shore}, {Sierpowska-Bartosik},
  {Sillanp{\"a}{\"a}}, {Sitarek}, {Sobczynska}, {Spanier}, {Spiro}, {Stamerra},
  {Steinke}, {Struebig}, {Suric}, {Takalo}, {Tavecchio}, {Temnikov}, {Terzic},
  {Tescaro}, {Teshima}, {Torres}, {Vankov}, {Wagner}, {Weitzel}, {Zabalza},
  {Zandanel}, {Zanin}, {Neronov}, \& {Semikoz}}]{magic}
{Aleksi{\'c}}, J., {Antonelli}, L.~A., {Antoranz}, P., {et~al.} 2010, \aap,
  524, A77, \dodoi{10.1051/0004-6361/201014747}

\bibitem[{Bai {et~al.}(2019)}]{lhaaso}
Bai, X., {et~al.} 2019.
\newblock \doarXiv{1905.02773}

\bibitem[{{Banerjee} \& {Jedamzik}(2004)}]{banerjee_jedamzik}
{Banerjee}, R., \& {Jedamzik}, K. 2004, \prd, 70, 123003,
  \dodoi{10.1103/PhysRevD.70.123003}

\bibitem[{Berezinsky \& Kalashev(2016)}]{Berezinsky:2016feh}
Berezinsky, V., \& Kalashev, O. 2016, Phys. Rev. D, 94, 023007,
  \dodoi{10.1103/PhysRevD.94.023007}

\bibitem[{{Bowman} {et~al.}(2018){Bowman}, {Rogers}, {Monsalve}, {Mozdzen}, \&
  {Mahesh}}]{edges2}
{Bowman}, J.~D., {Rogers}, A. E.~E., {Monsalve}, R.~A., {Mozdzen}, T.~J., \&
  {Mahesh}, N. 2018, \nat, 555, 67, \dodoi{10.1038/nature25792}

\bibitem[{{Budnev} {et~al.}(2020){Budnev}, {Astapov}, {Bezyazeekov}, {Bonvech},
  {Boreyko}, {Borodin}, {Br{\"u}ckner}, {Bulan}, {Chernov}, {Chernykh},
  {Chiavassa}, {Dyachok}, {Fedorov}, {Gafarov}, {Garmash}, {Grebenyuk},
  {Gress}, {Gress}, {Grinyuk}, {Grishin}, {Horns}, {Ivanova}, {Kalmykov},
  {Kazarina}, {Kindin}, {Kiryuhin}, {Kokoulin}, {Kompaniets}, {Kostunin},
  {Korosteleva}, {Kozhin}, {Kravchenko}, {Kryukov}, {Kuzmichev}, {Lagutin},
  {Lemeshev}, {Lubsand orzhiev}, {Lubsandorzhiev}, {Mirgazov}, {Mirzoyan},
  {Monkhoev}, {Osipova}, {Pakhorukov}, {Pan}, {Panasyuk}, {Pankov},
  {Podgrudkov}, {Poleschuk}, {Popesku}, {Popova}, {Porelli}, {Postnikov},
  {Prosin}, {Ptuskin}, {Petrukhin}, {Pushnin}, {Raikin}, {Rjabov}, {Rubtsov},
  {Sagan}, {Samoliga}, {Sidorenkov}, {Silaev}, {Silaev (junior}, {Skurikhin},
  {Slunecka}, {Sokolov}, {Sveshnikova}, {Suvorkin}, {Tabolenko}, {Tanaev},
  {Tarashansky}, {Ternovoy}, {Tkachenko}, {Tkachev}, {Tluczykont}, {Ushakov},
  {Vaidyanathan}, {Volchugov}, {Voronin}, {Wischnewski}, {Yashin},
  {Zagorodnikov}, \& {Zhurov}}]{taiga}
{Budnev}, N., {Astapov}, I., {Bezyazeekov}, P., {et~al.} 2020, Journal of
  Instrumentation, 15, C09031, \dodoi{10.1088/1748-0221/15/09/C09031}

\bibitem[{{Couti{\~n}o de Leon} {et~al.}(2019){Couti{\~n}o de Leon}, {Alonso},
  {Rosa-Gonzalez}, \& {Longinotti}}]{mrk_hawc}
{Couti{\~n}o de Leon}, S., {Alonso}, A.~C., {Rosa-Gonzalez}, D., \&
  {Longinotti}, A.~L. 2019, in International Cosmic Ray Conference, Vol.~36,
  36th International Cosmic Ray Conference (ICRC2019), 654.
\newblock \doarXiv{1909.01179}

\bibitem[{{Durrer} \& {Neronov}(2013)}]{durrer_neronov}
{Durrer}, R., \& {Neronov}, A. 2013, \aapr, 21, 62,
  \dodoi{10.1007/s00159-013-0062-7}

\bibitem[{{Fujita} \& {Kamada}(2016)}]{fujita16}
{Fujita}, T., \& {Kamada}, K. 2016, \prd, 93, 083520,
  \dodoi{10.1103/PhysRevD.93.083520}

\bibitem[{{Giacalone} \& {Jokipii}(1999)}]{jokipii99}
{Giacalone}, J., \& {Jokipii}, J.~R. 1999, \apj, 520, 204,
  \dodoi{10.1086/307452}

\bibitem[{Gilmore {et~al.}(2012)Gilmore, Somerville, Primack, \&
  Dominguez}]{Gilmore:2011ks}
Gilmore, R.~C., Somerville, R.~S., Primack, J.~R., \& Dominguez, A. 2012, Mon.
  Not. Roy. Astron. Soc., 422, 3189, \dodoi{10.1111/j.1365-2966.2012.20841.x}

\bibitem[{{Giovannini} \& {Shaposhnikov}(1998)}]{giovaninni97}
{Giovannini}, M., \& {Shaposhnikov}, M.~E. 1998, \prd, 57, 2186,
  \dodoi{10.1103/PhysRevD.57.2186}

\bibitem[{{Giroletti} {et~al.}(2008){Giroletti}, {Giovannini}, {Cotton},
  {Taylor}, {P{\'e}rez-Torres}, {Chiaberge}, \& {Edwards}}]{kpc_jet}
{Giroletti}, M., {Giovannini}, G., {Cotton}, W.~D., {et~al.} 2008, \aap, 488,
  905, \dodoi{10.1051/0004-6361:200809784}

\bibitem[{Holler {et~al.}(2018)Holler, Berge, Hahn, Khangulyan, \&
  Parsons}]{crab_ang_size}
Holler, M., Berge, D., Hahn, J., Khangulyan, D., \& Parsons, R.~D. 2018, PoS,
  ICRC2017, 676, \dodoi{10.22323/1.301.0676}

\bibitem[{Jasche \& Lavaux(2019)}]{Jasche2019}
Jasche, J., \& Lavaux, G. 2019, Astronomy \& Astrophysics, 625, A64,
  \dodoi{10.1051/0004-6361/201833710}

\bibitem[{Jasche \& Wandelt(2013)}]{borg_jw_13}
Jasche, J., \& Wandelt, B.~D. 2013, Monthly Notices of the Royal Astronomical
  Society, 432, 894, \dodoi{10.1093/mnras/stt449}

\bibitem[{Jedamzik \& Pogosian(2020)}]{jedamzik_pogosyan}
Jedamzik, K., \& Pogosian, L. 2020, Phys. Rev. Lett., 125, 181302,
  \dodoi{10.1103/PhysRevLett.125.181302}

\bibitem[{{Jedamzik} \& {Saveliev}(2019)}]{jedamzik_saveliev}
{Jedamzik}, K., \& {Saveliev}, A. 2019, \prl, 123, 021301,
  \dodoi{10.1103/PhysRevLett.123.021301}

\bibitem[{Kachelriess {et~al.}(2012)Kachelriess, Ostapchenko, \&
  Tomas}]{Kachelriess:2011bi}
Kachelriess, M., Ostapchenko, S., \& Tomas, R. 2012, Comput. Phys. Commun.,
  183, 1036, \dodoi{10.1016/j.cpc.2011.12.025}

\bibitem[{Kalashev \& Kido(2015)}]{Kalashev:2014xna}
Kalashev, O., \& Kido, E. 2015, J. Exp. Theor. Phys., 120, 790,
  \dodoi{10.1134/S1063776115040056}

\bibitem[{Kerszberg {et~al.}(2017)Kerszberg, Kraus, \&
  Kolitzus~D.}]{hess_icrc2017}
Kerszberg, D., Kraus, M., \& Kolitzus~D., e.~a. 2017

\bibitem[{{Kronberg}(1994)}]{kronberg94}
{Kronberg}, P.~P. 1994, Reports on Progress in Physics, 57, 325,
  \dodoi{10.1088/0034-4885/57/4/001}

\bibitem[{Lavaux \& Hudson(2011)}]{Lavaux2011_TMPP}
Lavaux, G., \& Hudson, M.~J. 2011, Monthly Notices of the Royal Astronomical
  Society, 416, 2840, \dodoi{10.1111/j.1365-2966.2011.19233.x}

\bibitem[{{Marinacci} {et~al.}(2018){Marinacci}, {Vogelsberger}, {Pakmor},
  {Torrey}, {Springel}, {Hernquist}, {Nelson}, {Weinberger}, {Pillepich},
  {Naiman}, \& {Genel}}]{illistristng}
{Marinacci}, F., {Vogelsberger}, M., {Pakmor}, R., {et~al.} 2018, \mnras, 480,
  5113, \dodoi{10.1093/mnras/sty2206}

\bibitem[{Natwariya \& Bhatt(2020)}]{edges1}
Natwariya, P.~K., \& Bhatt, J.~R. 2020, Mon. Not. Roy. Astron. Soc., 497, L35,
  \dodoi{10.1093/mnrasl/slaa108}

\bibitem[{{Neronov} \& {Semikoz}(2009)}]{neronov_semikoz09}
{Neronov}, A., \& {Semikoz}, D.~V. 2009, \prd, 80, 123012,
  \dodoi{10.1103/PhysRevD.80.123012}

\bibitem[{{Neronov} \& {Vovk}(2010)}]{neronov_vovk}
{Neronov}, A., \& {Vovk}, I. 2010, Science, 328, 73,
  \dodoi{10.1126/science.1184192}

\bibitem[{{Planck Collaboration}(2016)}]{planck16}
{Planck Collaboration}. 2016, \aap, 596, A103,
  \dodoi{10.1051/0004-6361/201528033}

\bibitem[{{Riess} {et~al.}(2019){Riess}, {Casertano}, {Yuan}, {Macri}, \&
  {Scolnic}}]{sn_H0}
{Riess}, A.~G., {Casertano}, S., {Yuan}, W., {Macri}, L.~M., \& {Scolnic}, D.
  2019, \apj, 876, 85, \dodoi{10.3847/1538-4357/ab1422}

\bibitem[{Taylor {et~al.}(2011)Taylor, Vovk, \& Neronov}]{Taylor:2011bn}
Taylor, A., Vovk, I., \& Neronov, A. 2011, Astron. Astrophys., 529, A144,
  \dodoi{10.1051/0004-6361/201116441}

\bibitem[{{Wong} {et~al.}(2020)}]{lensing_H0}
{Wong}, K.~C., {et~al.} 2020, \mnras, \dodoi{10.1093/mnras/stz3094}

\end{thebibliography}
\bibliographystyle{aasjournal}

\end{document}